\documentclass[10pt,letterpaper,twocolumn]{article} 

\usepackage{ol2}
\usepackage[utf8]{inputenc}
\usepackage{hyperref}
\usepackage{epsfig}
\usepackage{graphicx}
\usepackage{amsmath}
\usepackage{amssymb}
\usepackage{ulem}
\usepackage{color}

\graphicspath{{images/}}

\begin{document}

\twocolumn[
\title{Video-rate laser Doppler vibrometry by heterodyne holography}
\author{Benjamin Samson$^1$, Fr\'ed\'eric Verpillat$^2$, Michel Gross$^3$ and Michael Atlan$^1$}
\affiliation{
$^1$ Institut Langevin. Fondation Pierre-Gilles de Gennes. Centre National de la Recherche Scientifique (CNRS) UMR 7587, Institut National de la Sant\'e et de la Recherche M\'edicale (INSERM) U 979, Universit\'e Pierre et Marie Curie (UPMC), Universit\'e Paris 7. \'Ecole Sup\'erieure de Physique et de Chimie Industrielles - 10 rue Vauquelin. 75005 Paris. France\\
$^2$ Laboratoire Kastler-Brossel. \'Ecole Normale
Sup\'erieure. CNRS UMR 8552, UPMC, 24 rue Lhomond 75005 Paris. France\\
$^3$ Laboratoire Charles Coulomb. CNRS UMR 5221, Universit\'e Montpellier 2. Place E. Bataillon, 34000 Montpellier. France
}
\date{}
\begin{abstract}
We report a demonstration video-rate heterodyne holography in off-axis configuration. Reconstruction and display of 1 Megapixel holograms is achieved at 24 frames per second, with a graphics processing unit. Our claims are validated with real-time screening of steady-state vibration amplitudes in a wide-field, non-contact vibrometry experiment.
\end{abstract}
\maketitle
]

The laser Doppler method is the most common optical interferometry technique used for non-contact measurements of mechanical vibrations. Though highly effective for single-point vibration analysis, this technique is much less adapted to wide-field imaging than holography. Homodyne \cite{Powell1965, Picart2003, Pedrini2006} and heterodyne \cite{Aleksoff1971, JoudLaloe2009, JoudVerpillat2009} holographic recordings in off-axis configuration enabled reliable measurements of mechanical vibrations, but none of them allowed real-time monitoring, which is an essential feature. Matching the display rate of optically-measured Megapixel digital holograms with real-time imaging standards is demanding in terms of computational power. Holographic measurements are performed in a diffraction plane. Hence, image formation requires to simulate the back-propagation of an optical field. Such propagation involves turning the data measured in the plane of an array detector into a reciprocal plane with at least one bi-dimensional numerical Fourier transformation, typically a Fast-Fourier transform (FFT). Recently, real-time display of digital holograms with Graphics Processing Units (GPUs) \cite{ShimobabaSato2008, Ahrenberg2009} has alleviated the issue of the high computational workload needed for such image reconstruction. Parallel computations on GPU consistently increase the throughput with respect to CPU for computer-generated holograms, which demonstrated the performance of GPUs in streamline image processing \cite{AhrenbergBenzie2006}, \cite{MasudaIto2006}.

In this letter, we report an experimental demonstration of an image acquisition scheme designed to perform video-rate image reconstruction and display from heterodyne holographic measurements on a 1 Megapixel sensor array. Image reconstruction of steady-state vibration modes up to 100 kHz at a rate of 24 images per second is achieved. GPU processing is shown to enable holographic reconstruction and display with 3 FFT calculations per recorded frame, which covers the processing throughput needs of three reconstruction approaches : the \textit{convolution}, \textit{angular spectrum}, and \textit{Fresnel transform} methods \cite{KimYuMann2006}.

\begin{figure}[]
\centering
\includegraphics[width = 8.0 cm]{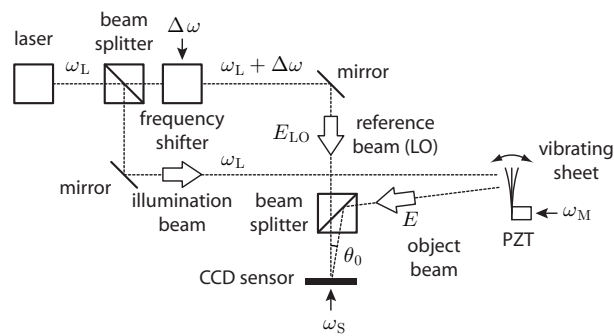}
\caption{Experimental image acquisition setup.}
\label{fig_setup}
\end{figure}

\begin{figure}[]
\centering
\includegraphics[width = 7.0 cm]{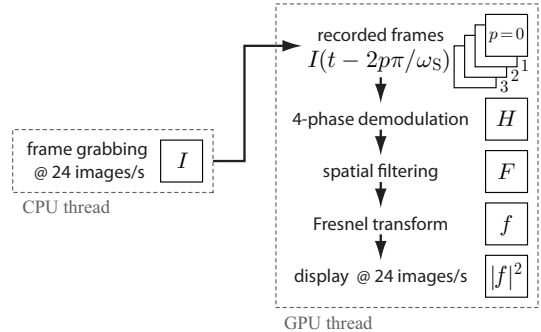}
\caption{Algorithmic layout of holographic rendering.}
\label{fig_algo}
\end{figure}

The acquisition setup consists of an off-axis, frequency-shifting holographic scheme, used to perform a multipixel heterodyne detection of optical modulation sidebands. Optical heterodyning is a process for placing information at frequencies of interest (e.g. the mechanical vibration of an object under investigation) into a useful frequency range by mixing the frequency content of the probe beam with a reference (or local oscillator, LO) beam. The optical frequency of the reference beam is shifted to generate a beat frequency of the interference pattern within the sensor bandwidth, which carries the information at the original frequency of interest. The Mach-Zehnder heterodyne interferometer used for the detection of an object field $E$ in reflective geometry, beating against a LO field $E_{\rm LO}$, is sketched in fig. \ref{fig_setup}.
The main optical radiation field is provided by a 100 mW, single-mode, doubled Nd:YAG laser (Oxxius SLIM 532) at wavelength $\lambda = 532$ nm, and optical frequency $\omega_{\rm L} / (2 \pi) = 5.6 \times 10^{14} \, \rm Hz$. The optical frequency of the LO beam is shifted by an arbitrary quantity $\Delta \omega$ in the low radiofrequency (RF) range ($0 \, {\rm Hz} \leq \Delta \omega / (2 \pi) \leq 100 \, {\rm kHz}$) by two acousto-optic modulators (AA-electronics, MT80-A1.5-VIS). The LO field has the form $E_{\rm LO} = {\cal E}_{\rm LO} {\rm e} ^{i (\omega_{\rm L} + \Delta \omega) t}$, where $\cal E_{\rm LO}$ is its complex amplitude. The object studied is a sheet of paper, whose lateral dimensions are $9 \times 26 \, \rm mm$, shined over $9 \times 17 \, \rm mm$. It is attached to a piezo-electric transducer (PZT, Thorlabs AE0505D08F), vibrating sinusoidally, driven at 10 V. Its local vibration is $z(t) = z_{\rm max}  \sin(\omega_{\rm M} t)$, where $z_{\rm max}$ is the vibration amplitude, provokes a modulation of the optical path length of the object field. It induces a local optical phase modulation of the backscattered field at the mechanical vibration frequency $\phi(t) = \phi_{0} \sin(\omega_{\rm M} t)$, where $\phi_{0} = 4 \pi z_{\rm max} / \lambda$ is the modulation depth of the optical phase. Holographic analysis of objects in sinusoidal vibration with a frequency-shifted LO beam tuned to the $n$-th optical modulation sideband was carried-out extensively in \cite{Aleksoff1971}. The frequency filtering properties of time-averaged holography were introduced in \cite{Powell1965}. The same filtering properties were described in digital holography \cite{PicartLeval2003, AtlanGross2007JOSAA}. It was shown that holograms of optical sidebands could be recorded selectively by matching the frequency shift of the local oscillator beam with the frequency of the band of interest. In the case of sinusoidal phase modulation, the hologram amplitude is of the form $J_n (\phi_{0})$, where $J_n$ is $n$-th order Bessel function of first kind. This modulation yields fringes that correspond to local extrema of $J_n$. It was also shown that imaging at harmonics of the vibration frequency could enable robust assessment of vibration amplitudes which are much greater than the optical wavelength \cite{JoudVerpillat2009}. The interference pattern $I$ is measured with a Pike F421-B camera on a Kodak KAI-04022 interline, progressive-scan CCD sensor ($2048 \times 2048$ pixels, pixel size $d_{\rm px} = 7.4 \, \mu \rm m$). The camera is run in binning mode (an effective pixel is made of 4 adjacent pixels); 16 bit, $1024 \times 1024$ pixels images are sampled at $\omega_{\rm S} / (2 \pi) = 24 \, \rm Hz$ throughout the experiments described hereafter. The RF command signals at frequencies $\Delta \omega$, $\omega_{\rm M}$, and  $\omega_{\rm S}$ are phase-locked. The temporal part of object field undergoing sinusoidal phase modulation can be decomposed in a basis of Bessel functions using the Jacobi–Anger identity
\begin{equation}
    E = {\cal E} {\rm e}^{ i \omega_{\rm L} t + i \phi(t)} = \sum_{n=-\infty}^{\infty} {\cal E}J_n(\phi_{0}) \, {\rm e}^{i( \omega_{\rm L} + n \omega_{\rm M})t}
\label{Eq_E}
\end{equation}
where ${\cal E}J_n(\phi_{0}) = {\cal E}_{n}$ is the weight of the optical modulation sideband of order $n$, and where $\cal E$ is the complex amplitude of the field. If the frequency detuning $\Delta \omega$ is set close to the n-th modulation harmonic, i.e. $\left| \Delta \omega - n \omega_{\rm M} \right| < \omega_{\rm S}$, and if the modulation frequency is much greater than the sampling frequency, i.e. $\omega_{\rm M} \gg \omega_{\rm S}$, the time-averaging-induced bandpass filter of the detection process will isolate the term of order $n$ in eq. \ref{Eq_E} and reject all other optical sidebands. In the sensor plane, the interference pattern of $E$ and $E_{\rm LO}$ takes the form $I(t) = \left| E + E_{\rm LO} \right|^2 = \left| E \right|^2 + \left| E_{\rm LO} \right|^2 + E E^*_{\rm LO} + E^* E_{\rm LO}$, where $^*$ denotes the complex conjugate. The frame $I(t)$ acquired at time $t$ by the framegrabber is moved to a frame buffer in the GPU RAM by a CPU thread (fig. \ref{fig_algo}).

To detect the heterodyne signal of interest $E E^*_{\rm LO} = {\cal E}^*_{\rm LO} {\cal E}_{n} {\rm e} ^{i (n\omega_{\rm M} - \Delta \omega) t}$, a sliding 4-phase temporal demodulation is performed. The intermediate frequency $n\omega_{\rm M} - \Delta \omega$ is set within the camera bandwidth to be sampled efficiently. More precisely, for a LO detuning $\Delta \omega = n\omega_{\rm M} - \omega_{\rm S}/4$, the modulation sideband ${\cal E}_{n}$ beats at the frequency $\omega_{\rm S}/4$ (6 Hz) in $I(t)$. To detect it, the following quantity is formed
\begin{equation}
    H(t) = \sum _{p=0}^{3} I(t-{2 p \pi / \omega_{\rm S}}) \, {\rm e}^{i p \pi /2}
\label{Eq_demod}
\end{equation}
$H$ is a complex-valued array, proportional to the diffracted field $E$ in the sensor plane. Its calculation requires the allocation of a stack of 4 arrays in the GPU memory, filled with $I(t)$, $I(t-{2 \pi / \omega_{\rm S}})$, $I(t-{4 \pi / \omega_{\rm S}})$, and $I(t-{6 \pi / \omega_{\rm S}})$. Each new frame grabbed at instant $t$ yields a shift of the stack (fig. \ref{fig_algo}) : $I(t)$ replaces the array $I(t-{8 \pi / \omega_{\rm S}})$.

In off-axis configuration, the spatial spectrum in the reciprocal plane ($k_x,k_y$) of the term $E E^*_{\rm LO}$ in the expression of $I(t)$ is shifted by the projection of the wave vector difference along a transverse direction $x$ in the sensor plane $\Delta k_x \sim 2 i \pi \theta_0 / \lambda$. Spatial filtering \cite{Cuche2000} of the time-demodulated signal $H(t)$ in off-axis geometry is used to remove the remaining contributions of the zero-order terms $|E|^2$ and $|E_{\rm LO}|^2$ and the twin-image term $E^* E_{\rm LO}$ to enhance the detection sensitivity. It is made by multiplying the ($k_x,k_y$) spectrum of $H$ by a mask $M$, allowing only frequencies in the neighborhood of $\Delta k$ to pass.
\begin{equation}
    F(t) = {\cal F}^{-1}\{  M {\cal F}\{ H(t) \} \}
\label{Eq_filtre}
\end{equation}
where ${\cal F}$ is a spatial FFT and ${\cal F}^{-1}$ is an inverse spatial FFT. This operation is handled by the GPU. Only the heterodyne contribution of interest $F(t) = {\cal E}^*_{\rm LO} {\cal E}_{n}$ remains in the filtered frame.

\begin{figure}[]
\centering
\includegraphics[width = 8.0 cm]{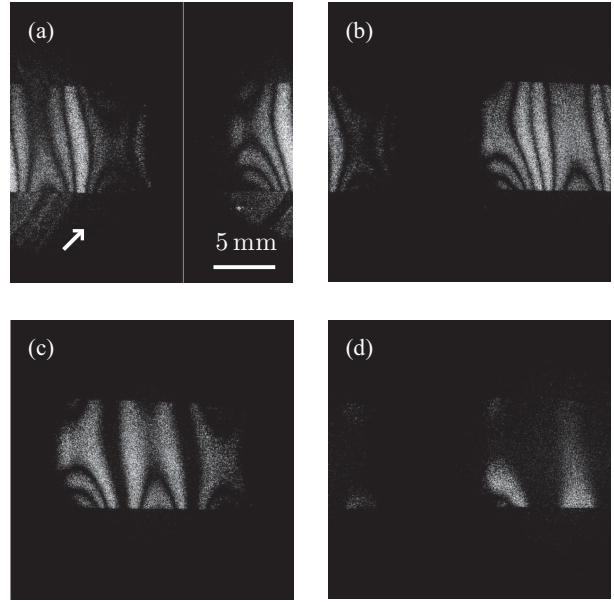}
\caption{
Time-averaged holographic maps of the $\omega_{\rm M} / (2 \pi) = 10 \, \rm kHz$ vibrational mode of the paper sheet for the modulation sidebands of order $n=0$ (a), $n=1$ (b), $n=3$ (c), and $n=7$ (d). $1024 \times 1024$ pixels renderings. The detection sideband sweep is reported in media 1. The first modulation sideband is screened for excitation frequencies swept from 0 Hz to 20 kHz in media 2 and from 0 Hz to 100 kHz in media 3. The effect of eq. \ref{Eq_filtre} filter is reported in media 4.}
\label{fig_maps}
\end{figure}

Image rendering from $F\propto {\cal E}_{n}$ in the sensor plane involves a scalar diffraction calculation in the Fresnel approximation, performed with a discrete Fresnel transform \cite{Schnars1994}. 
The hologram $f(t)$ back-propagated to the object plane is calculated by forming the FFT of the product of $F$ with a quadratic phase map, depending on the relative curvature of the wavefronts of $E$ and $E_{\rm LO}$ in the sensor plane $(x,y)$ via a distance parameter $\Delta z$. This calculation is handled by the GPU.
\begin{equation}
    f(t) =  {\cal F}\{ F(t) \, {\rm e}^ { i \pi (x^2 + y^2) / (\lambda \Delta z)} \}
\label{Eq_Fresnel}
\end{equation}
Finally, the GPU calculates the quantity $|f(t)|^2 \propto | J_n(4 \pi z_{\rm max} / \lambda) |^2$, which is a map in the object plane of the composition of the local vibration amplitude field $z_{\rm max}$ with the squared amplitude of the Bessel function of order $n$. Image brightness adjustment is
also performed by the GPU. Those maps are displayed in fig\ref{fig_maps} for the modulation sidebands of order $n=0$ (a), $n=1$ (b), $n=3$ (c), and $n=7$ (d). Excited at $\omega_{\rm M} / (2 \pi) = 10 \, \rm kHz$, the paper sheet builds up a steady-state vibrational mode with rectilinear nodes and bellies oriented along $y$, with a mechanical wavelength of $\sim$ 5 mm. In fig. \ref{fig_maps}(a), the non-moving support of the object in vibration is visible (arrow). A sweep of the detection sideband is reported in media 1. Additional vibrational patterns are screened for excitation frequencies swept from 0 Hz to 20 kHz (media 2) and from 0 Hz to 100 kHz (media 3), with a detection tuned to the first modulation sideband. The propensity of the filter of eq. \ref{Eq_filtre} to cancel-out spurious artefacts is assessed in real time in media 4.

The image reconstruction and display algorithm was elaborated with Microsoft Visual C++ 2008 integrated development environment and NVIDIA's Compute Unified Device Architecture (CUDA) software development kit 3.2. FFT calculations were made with the function \textit{cufftExecC2C()} from the CUFFT 3.2 library on single precision floating point arrays. The program was compiled and run on Microsoft Windows 7 - 64 bit. The computer hardware configuration was based on an ASUS P6T motherboard with a 2.67 GHz Intel core i7 920 CPU and a NVIDIA GeForce GTX 470 GPU. Image rendering calculations $I \to H \to F \to |f|^2$ are performed sequentially, in the main GPU thread (fig. \ref{fig_algo}); the whole processing time of one frame is reported in table \ref{tab_rendering}. For benchmark purposes, the rendering performance of $2048 \times 2048$ pixels recordings read out at $\omega_{\rm S} / (2 \pi) = 8 \, \rm Hz$ is also reported in table \ref{tab_rendering}.

\begin{table}
  \centering
  \caption{Benchmarks of image rendering time.}\begin{tabular}{cccc} \\ \hline
    \textbf{Array size (pixels)} & \textbf{time} \\ \hline
    $1024 \times 1024$ & 5.5 - 5.9 ms  \\
    $2048 \times 2048$  & 19.7 - 20.4 ms  \\ \hline
  \end{tabular}
\label{tab_rendering}
\end{table}
We have demonstrated that the detection and rendering of 1 Mega Pixel heterodyne holograms can be carried-out with a refreshment rate of 24 Hz with commodity computer graphics hardware. Video-rate optical monitoring of steady-state out-of-plane vibration amplitudes was reported. This demonstration opens the way to high throughput multipixel optical heterodyne sensing in real-time.

This work was funded by Fondation Pierre-Gilles de Gennes (FPGG014 grant), Agence Nationale de la Recherche (ANR-09-JCJC-0113 grant), And R\'egion \^Ile-de-France (C'Nano grant).


\end{document}